\documentclass[11pt,a4paper]{article}
\usepackage{amsmath,amssymb}
\usepackage{graphicx}
\usepackage{natbib}
\usepackage{hyperref}
\usepackage{lineno}
\usepackage[hmargin=3cm]{geometry}

\author{Nicolas Brantut$^1$ and L\'eo Petit$^2$\\
   $^{1}$ Department of Earth Sciences, University College London, London, UK\\
 $^{2}$ Laboratoire de G\'eologie, CNRS UMR 8538, \\D\'epartement de G\'eosciences, \'Ecole Normale Sup\'erieure, PSL University, Paris, France.}

\title{Micromechanics of rock damage and its recovery in cyclic loading conditions}

\date{\ }

\begin{document}

\maketitle

\begin{abstract}
  Under compressive stress, rock ``damage'' in the form of tensile microcracks is coupled to internal slip on microscopic interfaces, such as preexisting cracks and grain boundaries. In order to characterise the contribution of slip to the overall damage process, we conduct triaxial cyclic loading experiments on Westerly granite, and monitor volumetric strain and elastic wave velocity and anisotropy. Cyclic loading tests show large hysteresis in axial stress-strain behaviour that can be explained entirely by slip. Elastic wave velocity variations are observed only past a yield point, and show hysteresis with incomplete reversibility upon unloading. Irrecoverable volumetric strain and elastic wave velocity drop and anisotropy increase with increasing maximum stress, are amplified during hydrostatic decompression, and decrease logarithmically with time during hydrostatic hold periods after deformation cycles. The mechanical data and change in elastic properties are used to determine the proportion of mechanical work required to generate tensile cracks, which increases as the rock approaches failure but remains small, up to around 10\% of the net dissipated work per cycle. The pre-rupture deformation behaviour of rocks is qualitatively compatible with the mechanics of wing cracks. While tensile cracks are the source of large changes in rock physical properties, they are not systematically associated with significant energy dissipation and their aperture and growth is primarily controlled by friction, which exerts a dominant control on rock rheology in the brittle regime. Time-dependent friction along preexisting shear interfaces explains how tensile cracks can close under static conditions and produce recovery of elastic wave velocities over time. 
\end{abstract}

\section{Introduction}

Brittle deformation of rocks is characterised by macroscopic failure and fault slip, and is driven by frictional slip along preexisting interfaces and tensile fracturing at the microscale. Tensile microfracturing is what is generically termed ``damage'', and is a key geological marker of brittle deformation in the field. It has been extensively documented in regions surrounding major crustal faults, termed ``fault damage zones'', both in the field \citep[e.g.][and references therein]{faulkner10} and \textit{in situ} by seismological methods \citep[e.g.][]{li90,ben-zion91,li04,allam14,wang19}. Understanding how ``damage'' evolves in crustal rocks is crucial for our understanding of the seismic cycle, because the presence of tensile microcracks has a first order impact on both the post-yield rheology and the effective properties of rocks: it increases their porosity, permeability and their elastic compliance, and leads to anisotropy when the crack population has a preferred orientation \citep[e.g.][chap. 5]{paterson05}. In triaxial conditions, stress-induced tensile microcracks tend to be oriented parallel to the most compressive axis, which leads to vertically transverse isotropy in the crack orientation distribution and thus in elastic moduli.

Damage is often characterised and quantified by a measure of density and orientation of open microcracks: this is how fault damage zones have been documented in field and experimental studies \citep[e.g.][]{chester86,moore95,vermilye98,zang00,faulkner06,mitchell09,aben20b}. Under compressive nonhydrostatic stress states, open microcracks result from local tensile stresses between grains of different elastic properties, at the surface of open pores or at the edges of sliding defects such as grain boundaries \citep{tapponnier76,kranz83}. They are a consequence of shear deformation, but they contribute mostly to volumetric strain, by producing dilation \citep{brace66}, and to an increase in compliance perpendicular to their orientation. These dilational effects can be detected seismically \citep[e.g.,][]{gupta73,bonner74,lockner77a,schubnel03,aben19}. By contrast, microscale slip along preexisting flaws or grain boundaries produces additional shear compliance with hysteresis \citep[e.g.][]{kachanov82a}, as documented under uniaxial \citep{david12} and triaxial conditions \citep{david20}, but does not leave any direct microstructural evidence or seismic signature, and is therefore not usually considered as ``damage'' despite its major role in the micromechanics of brittle rock deformation.

The coupling between internal slip and tensile microcracking is the fundamental basis of our understanding of the rheology and physical properties of crystalline, non-porous rocks in the brittle regime under compressive loading. Microstructural evidence for such coupling was established by \citet{tapponnier76}, and many micromechanical models have successfully reproduced the key features of brittle deformation, including stress-strain behaviour, dilatancy, permeability and wave velocity evolution, based on the so-called ``wing crack'' model, whereby slip on shear cracks is coupled to tensile opening of ``wings'' at their tips \citep[e.g.][]{nemat-nasser82,horii85,nemat-nasser88, ashby90,basista98,simpson01,deshpande08,bhat11}. Spatial and temporal co-location of shear and tensile internal strain has been directly evidenced in laboratory tests by \citet{renard19}, giving support to the strong coupling arising from fracture mechanics models. Therefore, ``damage'' as measured in the field is expected to be associated to significant internal slip if it results from applied shear stresses. By contrast, tensile microfracturing can be generated pervasively under dynamic tensile loading, for instance near a fault during earthquake propagation \citep{doan09}. Thus, the partitioning of inelastic deformation between open cracks vs. shear cracks is a potential signature of the process at the origin of damage.

The state of damage at depth in the crust is however difficult to assess, due to (1) the strong load-path dependency and (2) the uncertainty regarding long-term persistence of both internal slip and tensile fracturing. External and internal stresses lead to closure or opening of existing (or newly formed) tensile microcracks, so that damage-induced property variations are partially reversible. A significant source of complexity is due to the microscale frictional rheology of shear cracks, which leads to hysteresis in stress-strain behaviour \citep{zoback75b} as well as in dilatancy \citep{scholz74,hadley76}, wave velocity \citep{holcomb81,passelegue18b} and permeability \citep{zoback75,mitchell08}. In addition, the time-dependency of friction leads to delayed crack closure, which results in time-dependent dilatancy and wave velocity recovery \citep{scholz74,holcomb81,brantut15b,meyer21}. It is likely that such time-dependent effects are responsible for at least part of the observed post-seismic variations in wave velocities around faults, which have been documented in many geological settings \citep[e.g.][]{li03,schaff04}.

In order to understand the impact and possible seismic signature of damage at depth in the crust, the respective contributions of tensile and shear cracks to the overall inelastic behaviour, as well as the reversibility and long-term evolution of the microstructural state, need to be quantified. 
Here, we follow the approach of \citet{holcomb81} and conduct cyclic loading tests in granite under triaxial compression, combining stress-strain measurements with wave velocity monitoring. A similar experimental setup was used recently by \citet{passelegue18b}, who documented in detail the reversibility of stress-induced anisotropy in granite. To go beyond the qualitative description of memory effects as originally reported by \citet{holcomb81}, we use wave velocities to compute effective moduli of the rock, which are sensitive to open cracks. By subtracting the contribution of those open cracks in the mechanical behaviour, we access separately to the contribution of internal slip. We estimate the partitioning of inelastic strain between microslip and opening as a function of proximity to failure (with increasing stress induced damage). In addition, we quantify the reversibility of dilatancy and elastic wave velocities upon unloading and as a function of time.


\section{Experimental methods}

\subsection{Mechanical testing}

Samples of Westerly granite were cyclically deformed in a triaxial apparatus \citep{eccles05} at confining pressures of 40, 80 and 120~MPa and with increasing differential stress, under dry conditions. 
The confining medium was silicone oil, and the confining pressure was controlled within $\pm0.5$~MPa. Axial load was measured externally by a load cell, and corrected for seal friction as in \citet{david20}. Sample shortening was measured with external Linear Variable Differential Transformers (LVDTs), corrected from elastic distortion of the loading column. Stresses and strains are taken positive in compression. The samples were instrumented with two pairs of axial and radial strain gauges, and jacketed in a nitrile sleeve fitted with an array of 14 P- and 2 Sh-sensitive piezoelectric transducers in direct contact with the rock surface. At regular time intervals throughout the tests, the transducer array was used to measure time of flight of P and Sh elastic waves at 4 different orientations with respect to the compression axis. Only direct arrivals are considered, and converted phases are not analysed. Details of the experimental technique and data processing can be found in \citet{brantut15b}.

The samples were first deformed in load-unload cycles with maximum differential stress gradually increasing by around 100~MPa between successive cycles. In each cycle, a constant axial deformation rate of $10^{-5}$~s$^{-1}$ was imposed until the target differential stress was reached. The samples were then unloaded at the same deformation rate down to minimum differential stress of around 20~MPa. These deformation cycles were conducted at increasing differential stress up to a maximum that was lower than the peak stress (failure strength) of the rock. Then, a second sequence of loading cycles was imposed, with maximum differential stress gradually decreasing by around 200~MPa between successive cycles.

An additional experiment was performed in a similar fashion, except that the load was fully removed after each unloading step, and the sample was held under constant hydrostatic pressure during a 24 hour period before the next loading step.

None of our samples were deformed until macroscopic failure. In order to normalise the stress levels used in our tests, we used an estimate of failure stress $Q_\mathrm{peak}$ appropriate for Westerly granite from \citet[][his Equation 7]{lockner98}:
\begin{linenomath}
  \begin{equation}
    Q_\mathrm{peak} (\text{MPa})= -8.3 + \sqrt{46 660.4 + 5 128.7\times P_\mathrm{c}}, \label{eq:Qpeak}
  \end{equation}
\end{linenomath}
where $P_\mathrm{c}$ is the confining pressure in MPa. This estimate is likely slightly too low for our specific samples, as evidenced by stress levels being sometimes greater than $Q_\mathrm{peak}$ even if samples did not experience failure. This discrepancy likely originates from natural sample variability, and is not critical to our interpretations since the main pressure dependency of strength is captured by Equation \eqref{eq:Qpeak}.

\subsection{Dynamic moduli inversion}

During triaxial compression, the elastic properties of rocks become strongly anisotropic, typically developing vertically transverse isotropy \citep[e.g.][]{schubnel03,passelegue18b}. In anisotropic materials, the times of flight obtained by pulse transmission between transducer pairs located at an angle from the plane of isotropy do not purely correspond to pure P or S phases. Assuming homogeneity, the wave velocities computed from such times of flight are \emph{group} velocities \citep[e.g.][]{thomsen86}. The computation of the stiffness matrix from group wave velocities is a nonlinear inverse problem where unknowns also include the phase angles between transducer pairs \citep{kovalyshen17}. We solve this problem in the least-square sense using the quasi-Newton algorithm \citet[][section 3.4.4]{tarantola05}, taking advantage of automatic differentiation for the computation of the jacobian matrix required at each iteration.


\section{Hysteresis in strain and elastic wave velocities}


\begin{figure}
  \centering
  \includegraphics{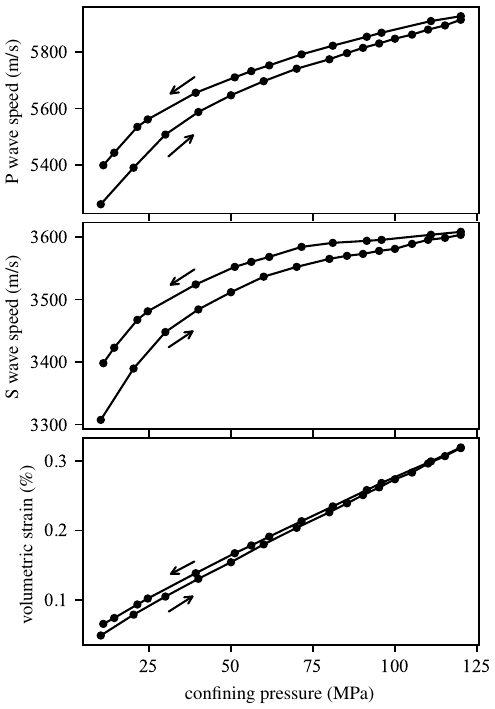}
  \caption{Average P and S wave velocities and volumetric strain during a hydrostatic pressure cycle.}
  \label{fig:hydrostat}
\end{figure}

Under purely hydrostatic conditions, Westerly granite exhibits some hysteretical behaviour in terms of elastic wave velocities and volumetric strain (Figure \ref{fig:hydrostat}): upon a full pressurisation-depressurisation cycle between 10 and 120~MPa, a permanent increase in $V_\mathrm{P}$ and $V_\mathrm{S}$ of around 3\% is observed, together with a small permanent decrease in bulk volume, of the order of 0.02\%. The hysteresis in wave velocity and strain indicates that some irreversible microcrack closure occurs during hydrostatic loading \citep{walsh65}.

\begin{figure}
  \centering
  \includegraphics{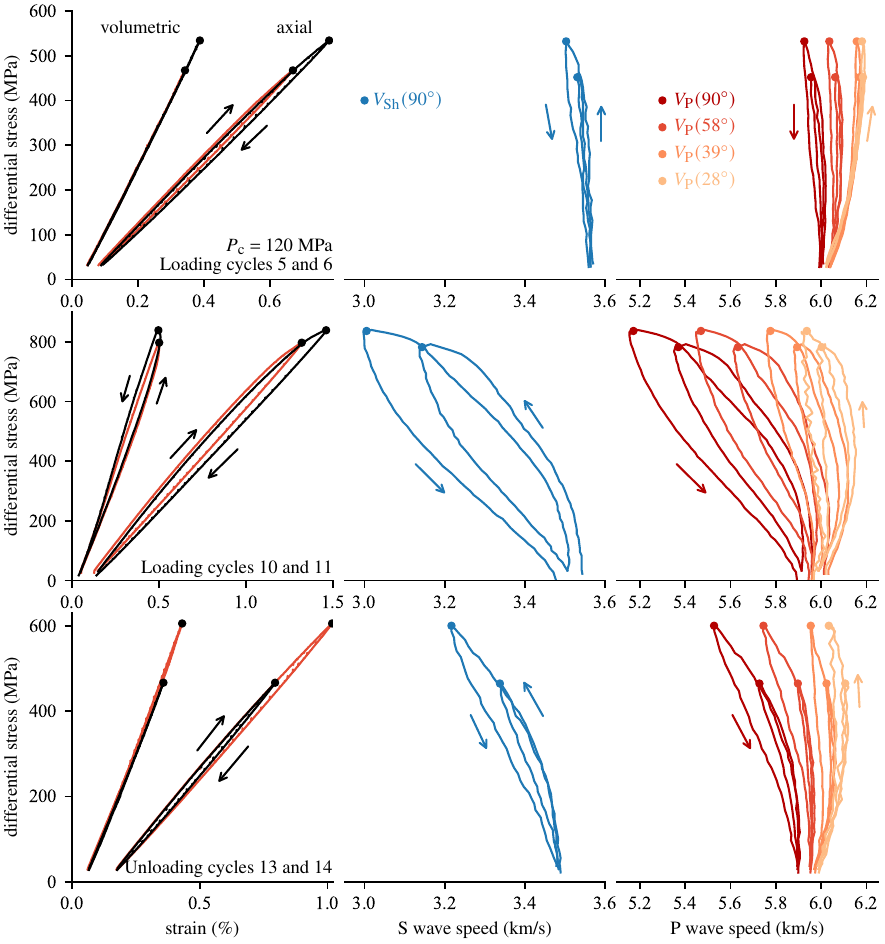}
  \caption{Evolution of differential stress, axial and volumetric strain, and P and S wave velocities during representative pairs of consecutive load-unload cycles in Westerly granite at 120~MPa confining pressure. The estimated peak stress at this pressure is (Equation \ref{eq:Qpeak}) 805~MPa. In left panels, red curves mark the stress-strain behaviour during the first of the two cycles. Legends indicate angles of propagation with respect to axial direction; $90^\circ$ is perpendicular to loading axis (horizontal), and $0^\circ$ is along loading axis (vertical).}
  \label{fig:typicalcycles}
\end{figure}

Under triaxial stress conditions, three main regimes can be distinguished depending on the maximum stress achieved during the cycle and the loading history (Figure \ref{fig:typicalcycles}). In successive loading cycles at relatively low stress conditions (typically below the onset of inelastic dilation), significant hysteresis was observed in differential stress vs. axial strain behaviour, whereas volumetric strain was essentially fully reversible. The elastic wave velocities varied only slightly: with increasing differential stress, horizontal P and S wave velocities remained approximately constant and a slight decrease initiated at the highest loads. Subvertical P wave velocity increased at low differential stress. Some hysteresis also occurs in wave velocities, with values remaining lower during unloading than during loading. All the measured quantities (stress, strain, wave velocities) at the maximum differential stress achieved in a given cycle were matched again
when stress returned to the same level in the subsequent cycle.

During loading cycles conducted to higher maximum stress, beyond the onset of measurable dilatancy, hysteresis was observed in both axial and volumetric strain. The horizontal wave velocities evolved in three steps: they remained approximately constant up to a stress threshold (around 300~MPa in the test at $P_\mathrm{c}=120$~MPa, independently from the cycles), then decreased linearly up to the previous maximum stress experienced by the rock, and then decreased significantly more beyond that stress. Wave velocities changed only very slightly at the onset of unloading, and returned gradually to their initial values with further unloading. The wave velocity recovery was almost total after unloading (as documented by \citet{passelegue18b}), except for a small permanent decrease. Upon reloading after a given cycle, the axial and volumetric strain as well as the wave velocities followed a qualitatively similar change compared to that in the previous cycle, but did not overlap until the previous maximum stress is reached.

When loading cycles were conducted to a lower maximum differential stress compared to overall maximum achieved during the experiment, the axial strain showed hysteresis, and was fully reversible when load was removed (Figure \ref{fig:typicalcycles}, bottom left). The volumetric strain showed some nonlinearity with stress but no hysteresis. Similarly to the pre-dilatancy behaviour, horizontal wave velocities tended to decrease with increasing load. As documented by \citet{holcomb81}, axial and volumetric strain, and wave velocities completely overlapped during reloading as long as the maximum stress did not exceed the overall maximum achieved during the test.

\begin{figure}
  \centering
  \includegraphics{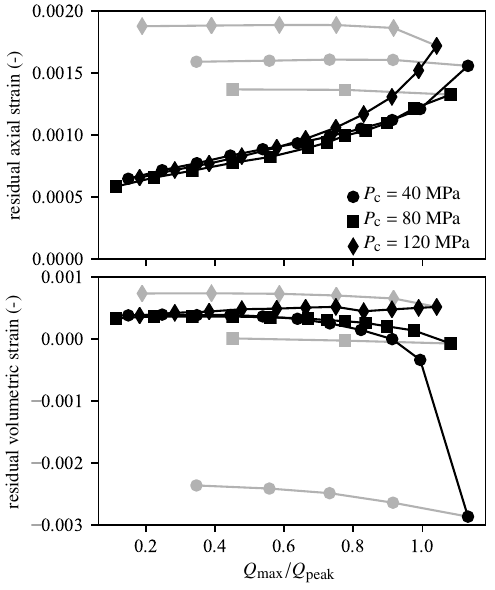}
  \caption{Irrecoverable axial and volumetric strain after unloading as a function of maximum differential stress achieved during each cycle, $Q_\mathrm{max}$, normalised by the estimated failure stress $Q_\mathrm{peak}$ (Equation \ref{eq:Qpeak}). Black symbols: loading cycles conducted at increasing maximum differential stresses. Grey symbols: loading cycles conducted at decreasing maximum differential stress.}
  \label{fig:residualstrain}
\end{figure}

The stress-strain behaviour and elastic wave velocities of Westerly granite measured here are entirely consistent with previous observations by \citet{zoback75b,holcomb81,passelegue18b}. Turning our attention to the state of the rock after each unloading stage, we observe a gradual increase in irrecoverable axial and volumetric strain with increasing maximum differential stress (Figure \ref{fig:residualstrain}). Irrecoverable dilatant volumetric strain was markedly larger at 40 MPa confining pressure, whereas it remained negligible at 120~MPa. By contrast, more axial permanent strain remained after unloading at high confining pressure compared to low confining pressure. In subsequent cycles conducted to lower maximum stress, permanent strain remained approximately constant.


\begin{figure}
  \centering
  \includegraphics{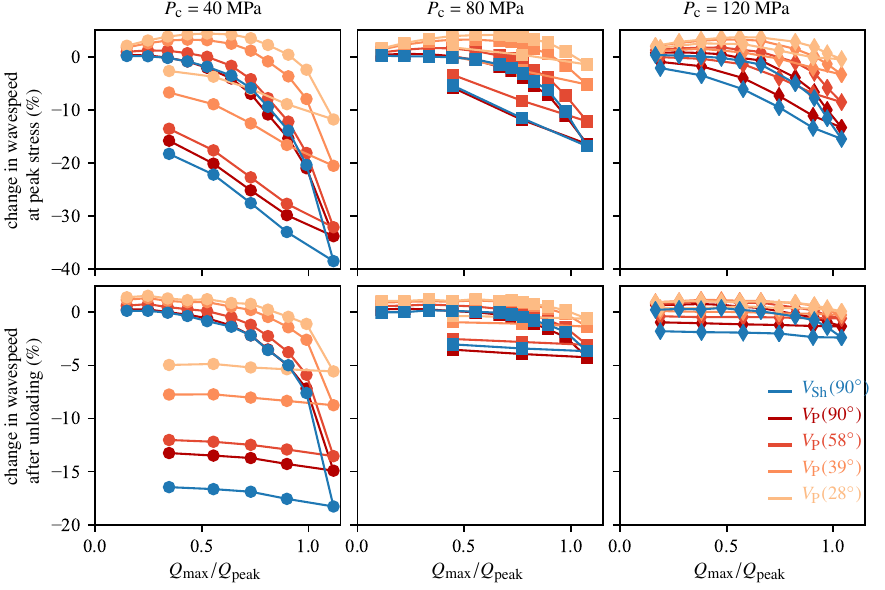}
  \caption{Relative change in elastic wave velocities at the maximum differential stress (top) and after unloading (bottom).}
  \label{fig:residualandmaxV}
\end{figure}

\begin{figure}
  \centering
  \includegraphics{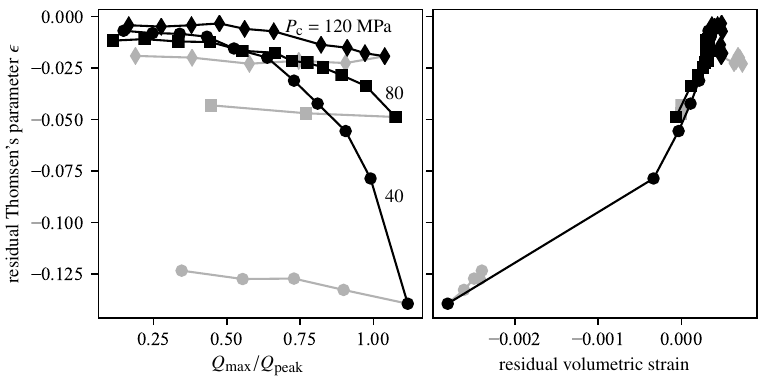}
  \caption{Thomsen's $\epsilon$ parameter measured after unloading as a function of (a) maximum differential stress achieved during each cycle, and (b) irrecoverable volumetric strain. Black symbols: loading cycles conducted at increasing maximum differential stresses. Grey symbols: loading cycles conducted at decreasing maximum differential stress.}
  \label{fig:residualthomsen}
\end{figure}

Elastic wave velocities decreased dramatically with increasing differential stress and recovered during unloading, but not always completely (Figure \ref{fig:residualandmaxV}). Permanent changes remained after cycles where the maximum differential stress exceeded about 50\% of the estimated reference peak stress, which roughly coincides with the stress threshold for dilatancy (even though such a quantity is difficult to quantify precisely, as stated by \citet[][p. 22]{hadley75}). The permanent drop in wave velocity increased with increasing maximum differential stress, and remained constant in any subsequent cycles where the previously achieved maximum differential stress was not overcome. Except for one cycle at high differential stress at 40~MPa confining pressure, residual wave velocity drops were typically of the order of a few percent. The elastic anisotropy that developed at high stress weakened but also remained nonzero after unloading (keeping in mind the caveat that unloading was not conducted down to zero but to about 20~MPa differential stress), with subvertical P wave velocities being higher than subhorizontal ones. The residual P wave anisotropy can be characterised by Thomsen's $\epsilon$ parameter defined as $\epsilon=(C_{11}-C_{33})/(2C_{33})$ (where $C_{ij}$ denote components of the elastic tensor in Voigt's notation), which showed a decrease towards more negative values after stress cycles with increasing differential stress (Figure \ref{fig:residualthomsen}). At 40 MPa confining pressure, a relatively large permanent anisotropy remained ($\epsilon\approx-.125$), in conjunction with a dilatant residual volumetric strain. With increasing confining pressure, the residual anisotropy decreased, down to around $\epsilon=-0.025$ after the largest stress cycle at 120 MPa confining pressure. In all cases, the anisotropy did not change significantly with subsequent loading cycles when the maximum stress achieved was less than the overall maximum reached during the test. The residual anisotropy was correlated to the residual volumetric strain (Figure \ref{fig:residualthomsen}).

\begin{figure}
  \centering
  \includegraphics{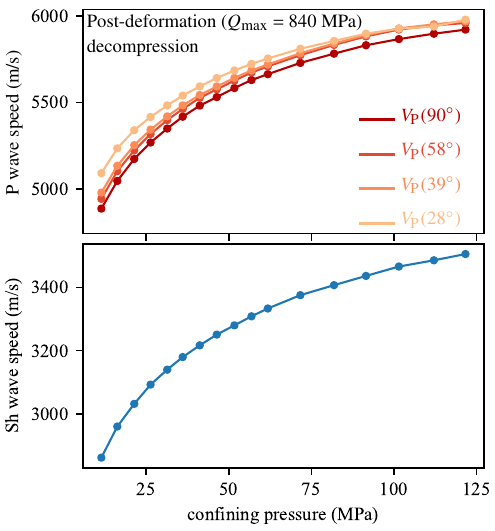}
  \caption{Change in elastic wave velocities during hydrostatic decompression after a series of 16 loading cycles up to 840~MPa differential stress.}
  \label{fig:decompress_vpvs}
\end{figure}

\begin{figure}
  \centering
  \includegraphics{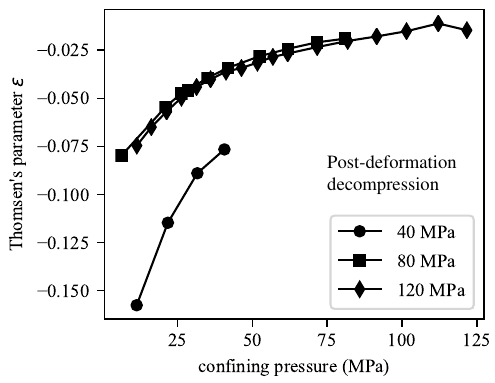}
  \caption{Evolution of Thomsen's parameter $\epsilon$ during hydrostatic decompression after deformation.}
  \label{fig:decompress_thomsen}
\end{figure}

After each experiment, elastic wave velocities decreased during hydrostatic decompression, down to values lower than those of the intact material (e.g., by about 10\% for the test conducted at 120 MPa confining pressure, cf. Figure \ref{fig:decompress_vpvs}). The decrease in elastic wave velocities was accompanied by a significant increase in anisotropy (Figure \ref{fig:decompress_thomsen}).




\section{Role of tensile microcracks}

\begin{figure}
  \centering
  \includegraphics{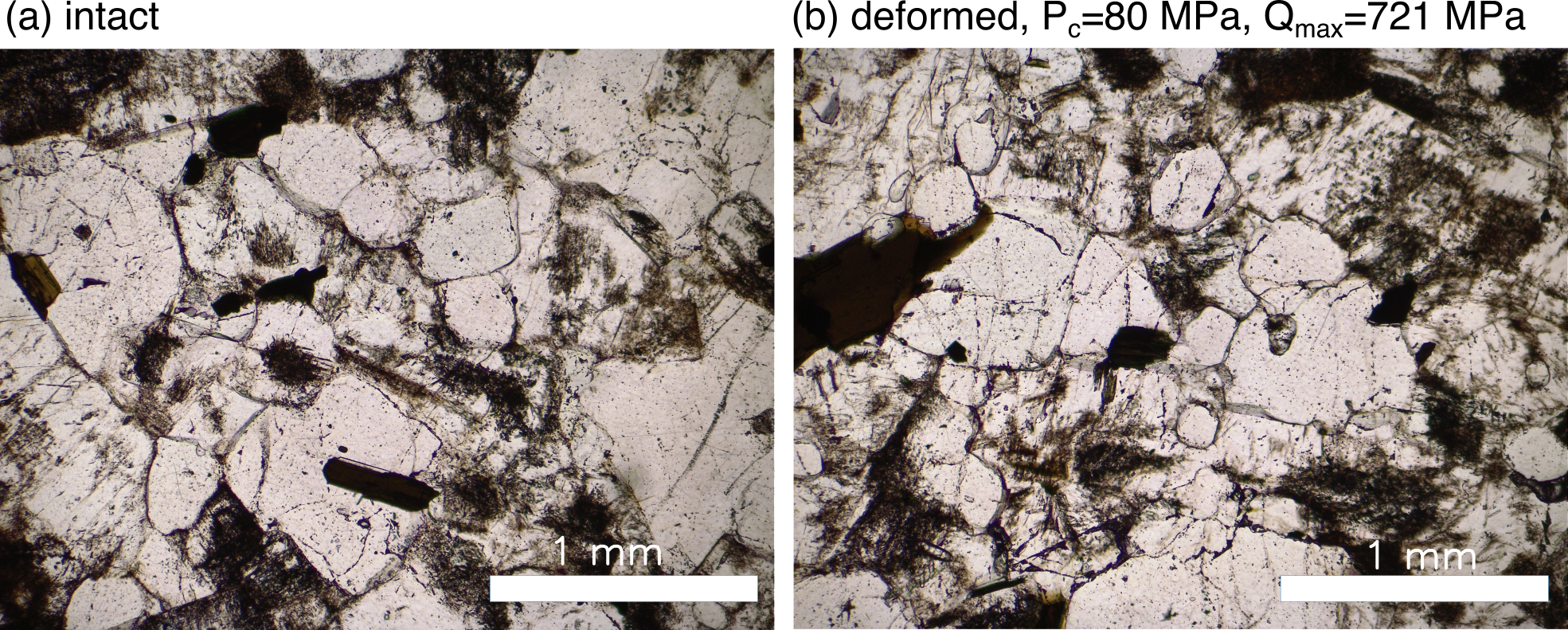}
  \caption{Optical micrographs of (a) intact and (b) deformed Westerly granite, in transmitted plane polarised light. Compression axis is vertical.}
  \label{fig:micro}
\end{figure}

Thin sections were made from an intact sample and from the sample deformed at 80~MPa confining pressure, which experienced a maximum differential stress of 721~MPa and a total of 14 load-unload cycles. Despite a post-deformation wave velocity at least 20\% lower than the intact rock, qualitative optical observations reveal very little, if any, signs of mode I intragranular microcrack damage beyond what is observed in the intact material (Figure \ref{fig:micro}). While it is well known that detection of stress-induced intragranular cracks is difficult with basic optical methods \citep{tapponnier76}, the apparent discrepancy between the large change in properties and the high stress level experienced by the sample and the lack of any widespread, visible microstructural evidence is quite striking. That being said, our observations are still consistent with the detailed microstructural work of \citet{tapponnier76}, who could only detect newly formed intragranular cracks at stresses significantly higher than the onset of dilatancy.

Therefore, it is likely that a significant source of dilatancy and wave velocity variations observed during the test, especially at low stress, was the opening of preexisting cracks or grain boundaries. The large recovery of properties during unloading implies that any newly open crack could be almost perfectly closable, i.e., crack faces should return into contact (at least partially) as load is removed.

Assuming that the anisotropic variations in elastic moduli arise from open microcracks, we use an effective medium model to invert the time of flight data for quantitative crack density parameters. Here we follow the approach of \citet{sayers95}, who consider an array of uniformly distributed penny-shaped microcracks embedded in a homogeneous, isotropic elastic matrix. Interactions between cracks are neglected, which is valid in the limit of small crack density, and crack orientations are assumed to be distributed in a transversely-isotropic manner, with symmetry axis along the most compressive load direction. \citet{sayers95} express the effective elastic properties of the cracked material in terms of the moduli of solid and five independent crack density parameters: two components $\alpha_{11}$ and $\alpha_{33}$ of a second-order crack density tensor, and three components $\beta_{1111}$, $\beta_{1133}$ and $\beta_{3333}$ of a fourth-order crack density tensor. The components $\alpha_{11}$ and $\alpha_{33}$ correspond to the vertical and horizontal crack densities defined by $N_\mathrm{v}c^3$, where $N_\mathrm{v}$ is the number of cracks per unit volume and $c$ is their average radius. The components $\beta_{ijkl}$ correspond to fourth order moments of the crack orientation distribution. Complete mathematical definitions can be found in \citet{sayers95}. The solid matrix Young's modulus is taken equal to $E=89$~GPa and its Poisson's ratio is $\nu=0.22$, chosen to match intact P and S wave velocities of $6.2$ and $3.7$~km/s, respectively. To better constrain the inversion, we use the prior knowledge that the $\beta$'s are a factor $\nu/2$ smaller than the $\alpha$'s \citep[][their equation 15]{sayers95}, i.e., roughly one order of magnitude smaller. The inverse problem is solved by the quasi-Newton method of \citet[][section 3.4.4]{tarantola05}, using the time of flights as observables and crack density parameters and phase angles as unknowns. Since we expect the $\alpha$'s to be of the order of $0.1$, we use a prior value of $0$ and a variance of $10^{-2}$ on all $\beta$'s.

\begin{figure}
  \centering
  \includegraphics{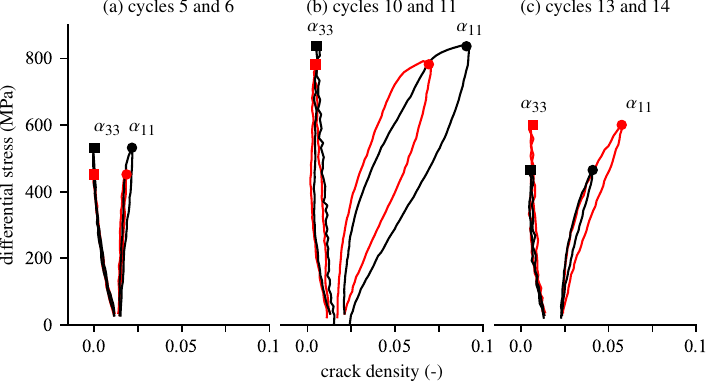}
  \caption{Evolution of vertical ($\alpha_{11}$) and  horizontal ($\alpha_{33}$) crack density with differential stress during representative pairs of consecutive load-unload cycles in Westerly granite at 120~MPa confining pressure. Red curves mark the behaviour during the first of the two cycles.}
  \label{fig:cyclescracks}
\end{figure}

The evolution of horizontal and vertical crack densities mirrors the evolution in elastic wave velocities (illustrated for representative cycles at 120~MPa confining pressure in Figure \ref{fig:cyclescracks}). In the low stress regime, the horizontal crack density decreases slightly with increasing stress, while the vertical crack density remains stable. Beyond the onset of dilatancy, the vertical crack density starts increasing and the horizontal crack density tends to stabilise. For cycles conducted up to relatively low differential stress, hysteresis is observed only in the vertical crack density, which is almost completely recoverable upon full unloading. At elevated differential stress, the vertical crack density increases with increasing stress, until the previous maximum stress is reached, and increases more markedly with further loading. Upon unloading, the vertical crack density remains high, and recovers only partially. The horizontal crack density remains more or less stable at high stress, and increases linearly with differential stress during unloading, showing hysteresis. In cycles conducted to lower maximum differential stresses compared to previous cycles, the evolution of crack density is perfectly reproducible during re-loading. No hysteresis is observed in horizontal crack density, but hysteresis remains in the evolution of vertical crack density.

Overall, the density of open, mostly vertical, microcracks evolve nonlinearly with stress, and decrease (i.e., open microcracks mechanically close) with strong hysteresis during unloading. Preexisting cracks make the material behave nonlinearly, i.e., compliance decreases with increasing stress. The formation of new cracks, either by extension or re-opening of preexisting ones, or by creation of new crack surfaces in intact material, is achieved only when the previously achieved maximum stress is overcome. When no new cracks are generated in a given load cycle, the crack density evolution during loading is reproducible, but hysteresis during unloading remains.

\begin{figure}
  \centering
  \includegraphics{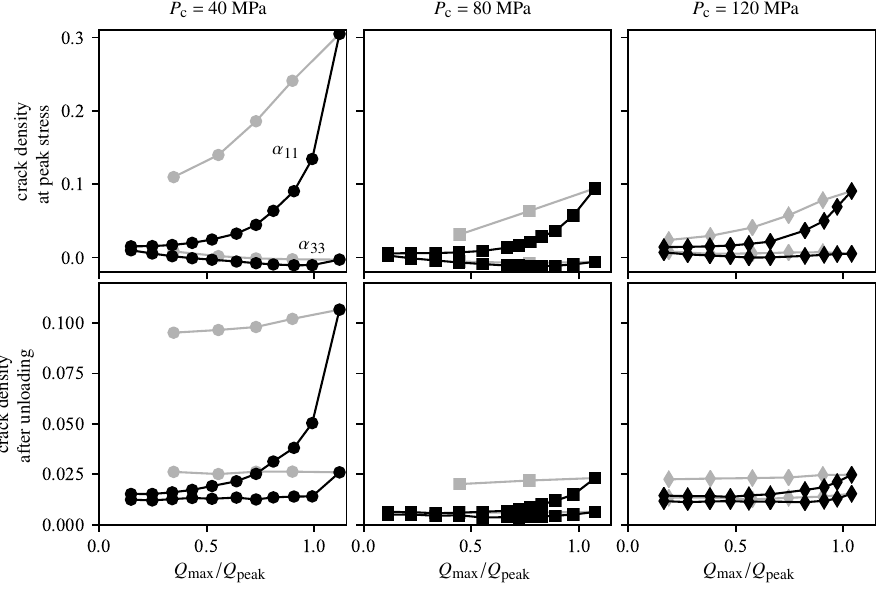}
  \caption{Vertical ($\alpha_{11}$) horizontal ($\alpha_{33}$) crack densities at the maximum differential stress (top) and after unloading (bottom) for each loading cycle. Black symbols: loading cycles conducted at increasing maximum differential stresses. Grey symbols: loading cycles conducted at decreasing maximum differential stress.}
  \label{fig:residualcracks}
\end{figure}

The residual crack density after unloading increases with increasing differential stress, in proportion to the maximum crack density reached during each cycle (Figure \ref{fig:residualcracks}). The maximum vertical crack density is of the order of 0.1 when differential stress approaches the rock strength at each confining pressure investigated; this value is consistent with the non-interaction approximation of \citeauthor{sayers95}'s effective medium model. The residual vertical crack density can be as high as 0.025 after unloading, and remains constant with any further cycles if the previously achieved maximum stress is not overcome.

\begin{figure}
  \centering
  \includegraphics{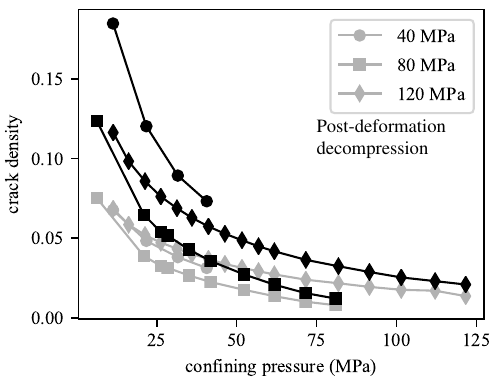}
  \caption{Vertical (black symbols) and horizontal (grey symbols) crack density during hydrostatic decompression after deformation.}
  \label{fig:decompress_cracks}
\end{figure}

During hydrostatic decompression, both vertical and horizontal crack densities increase with decreasing confining pressure, reaching values of the order of 0.1 at the lowest pressure (Figure \ref{fig:decompress_cracks}). The increase in anisotropy noted in Figure \ref{fig:decompress_thomsen} is explained by a stronger increase in vertical compared to horizontal crack density.


\section{Strain and energy partitioning: Inelastic vs. elastic effects of microcracks and microscale friction}
\label{sec:energy}

The combined knowledge of stress, strain and dynamic elastic moduli allows us to partially quantify the contributions of open microcracks and other microstructural changes in the total deformation of the rock. The total strain $\epsilon_{ij}^\mathrm{tot}$ in a homogeneous elementary volume of material is decomposed as \citep[][his equation 2.27]{rice75}
\begin{linenomath}\begin{equation}
  \epsilon_{ij}^\mathrm{tot} = (S^0_{ijkl}+\Delta S_{ijkl})\sigma_{kl} + \epsilon_{ij}^\mathrm{i},
\end{equation}\end{linenomath}
where $S^0_{ijkl}$ is the compliance tensor of the crack-free matrix, $\Delta S_{ijkl}$ is the change in compliance due to open cracks, $\sigma_{kl}$ is the stress tensor and $\epsilon^\mathrm{i}_{ij}$ is the strain associated with all inelastic processes, including internal sliding and crack growth.

\begin{figure}
  \centering
  \includegraphics{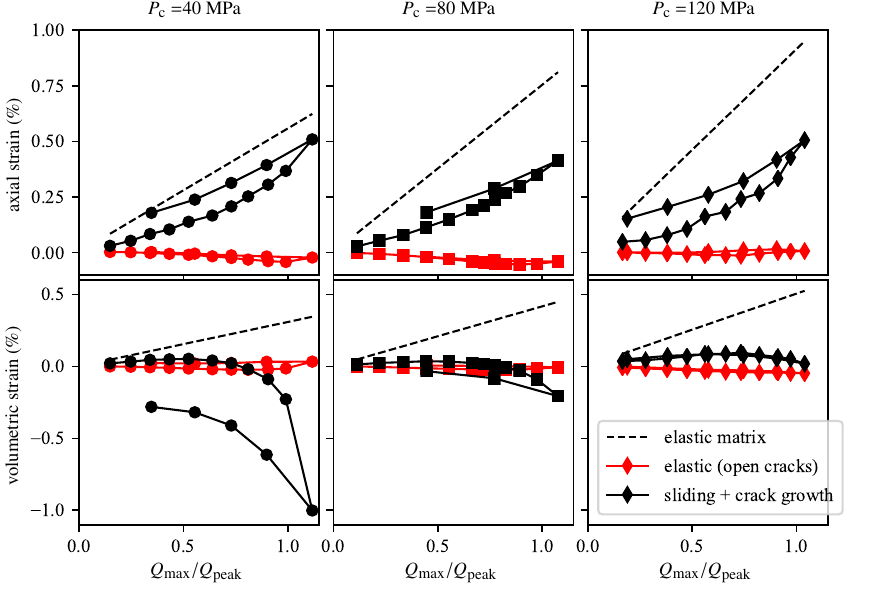}
  \caption{Axial (top) and volumetric (bottom) strain at the maximum differential stress of each cycle. Dashed lines are the elastic contributions of the solid, red curves at the elastic contributions of open microcracks, and black curves are the remaining inelastic contributions, which include internal slip and irreversible crack growth. All contributions sum to produce the observed stress-strain curves. The elastic contribution of open cracks is derived from inversion of wave velocities, and the contribution of slip and crack growth is derived by difference between measured strains and inverted elastic strains from other contributions.}
  \label{fig:strainpart}
\end{figure}

At all the tested confining pressures, the axial strain is dominated by the elastic contribution together with inelastic processes (Figure \ref{fig:strainpart}). The contribution of crack-induced compliance changes is practically negligible. In terms of volumetric strain, the elastic contribution is significant, producing additional compression, and the contribution of inelastic processes increases with increasing differential stress to produce dilation, which eventually dominates the total response. There again, purely elastic variations in compliance due to open cracks do not produce significant strains. In any case, the effect of open cracks is to increase elastic compliances in dry rocks, which would tend to produce \emph{more compressive strains} with increasing damage. The large dilatancy is therefore an effect of inelastic variations in crack length, sliding and crack opening.

While the crack-induced changes in compliance do not seem to produce much elastic strain by themselves, they produce inelastic strain and are associated with fracture energy dissipation or conversion into surface energy. The change in Gibbs' free energy per unit volume due to the growth of open cracks in an elastic medium is given by \citep[e.g.][, their equation 2.16]{kachanov12}
\begin{linenomath}\begin{equation}
  \delta G = \frac{1}{2}\delta S_{ijkl} \sigma_{ij}\sigma_{kl},
\end{equation}\end{linenomath}
where $S_{ijkl}$ is the compliance tensor, $\sigma$ is the stress tensor, and the symbol $\delta$ denotes a small increment of each quantity. Therefore, the integral
\begin{linenomath}\begin{equation} \label{eq:DG}
  \Delta G = \frac{1}{2}\int_S^{S+\Delta S} \delta S_{ijkl} \sigma_{ij}\sigma_{kl}
\end{equation}\end{linenomath}
corresponds to the total energy change associated with the increase in compliance due to the creation of new crack surfaces, which is thus equivalent to the fracture energy multiplied by the area of new cracks per unit volume of the material.

Throughout the loading cycles, the compliance keeps changing, decreasing during unloading and increasing during re-loading (Figure \ref{fig:typicalcycles}). However, not all these changes correspond to the formation of new cracks: the decrease in compliance during unloading is due to the formation of new contact points along preexisting cracks (no healing occurs), and the increase in compliance up to the previous maximum load is due to the re-opening of preexisting cracks. Therefore, only the compliance increase during the loading steps beyond any previously achieved maximum differential stress should be considered in the estimation of fracture energy by Equation \ref{eq:DG} (see path $BC$ in Figure \ref{fig:energysketch}). Despite this precaution, it is quite clear that more compliance variations would also be due to re-opening (i.e. removal of contact points) of preexisting cracks at all stages of the loading process, so that our estimate of fracture energy is an upper bound. An alternative method to obtain a less severe upper bound on integrated fracture energy per cycle from stress-strain curves is discussed in Appendix \ref{ax:energy}.

\begin{figure}
  \centering
  \includegraphics{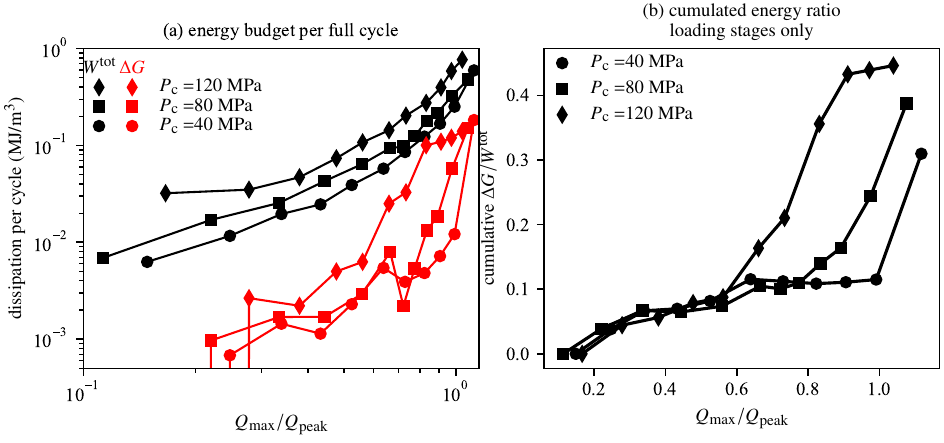}
  \caption{(a) Energy dissipation per cycle. Black curves correspond to the total dissipation, and red curves correspond to the contribution of microcrack growth computed from Equation \protect\eqref{eq:DG}, which is equivalent to the total fracture energy per unit volume. (b) Ratio of cumulated fracture energy over total dissipation per unit volume as a function of differential stress during loading stages only.}
  \label{fig:energypartall}
\end{figure}

We compute the energy change due to crack growth for each cycle of our experiments, and compare it to the total dissipation, given by
\begin{linenomath}\begin{equation}
  W^\mathrm{tot} = \int_C P_\mathrm{c}d\epsilon_{kk} + \int_C Qd\epsilon_\mathrm{ax},
\end{equation}\end{linenomath}
where $\epsilon_{kk}$ denotes the volumetric strain, $Q$ is the differential stress and $\epsilon_\mathrm{ax}$ is the axial strain. The symbol $\int_C$ denotes integration along the full strain cycle. Both the total dissipation and our estimate of crack-induced dissipation increase with increasing maximum differential stress for each cycle (Figure \ref{fig:energypartall}a). The contribution of crack growth per loading cycle is negligible up to relatively high stress, and grows rapidly near the failure stress. It remains typically around one order of magnitude smaller than the total dissipation. With increasing confining pressure, dissipation increases slightly and the data suggest that the contribution of cracks increases more gradually with increasing stress.

A significant part of the total dissipation by frictional processes occurs during the unloading and re-loading steps of each cycle, whereas crack growth occurs when load is increased beyond the previously achieved maximum (segment BC in Figure \ref{fig:energysketch}). The energy budget computed during only those loading steps indicates a growing contribution of fracturing with increasing stress, from around 10\% up to 40\% of the total dissipation near the peak stress (Figure \ref{fig:energypartall}b). At elevated confining pressure, the cumulated fracture energy (per unit rock volume) is a larger fraction of the total dissipation, whereas the inferred crack densities are comparable or even lower than at low pressure (Figure \ref{fig:residualcracks}). This indicates that deformation promotes the growth of fewer, larger cracks at low pressure and more numerous, shorter cracks at elevated pressure.

\section{Time dependency and recovery}

\begin{figure}
  \centering
  \includegraphics{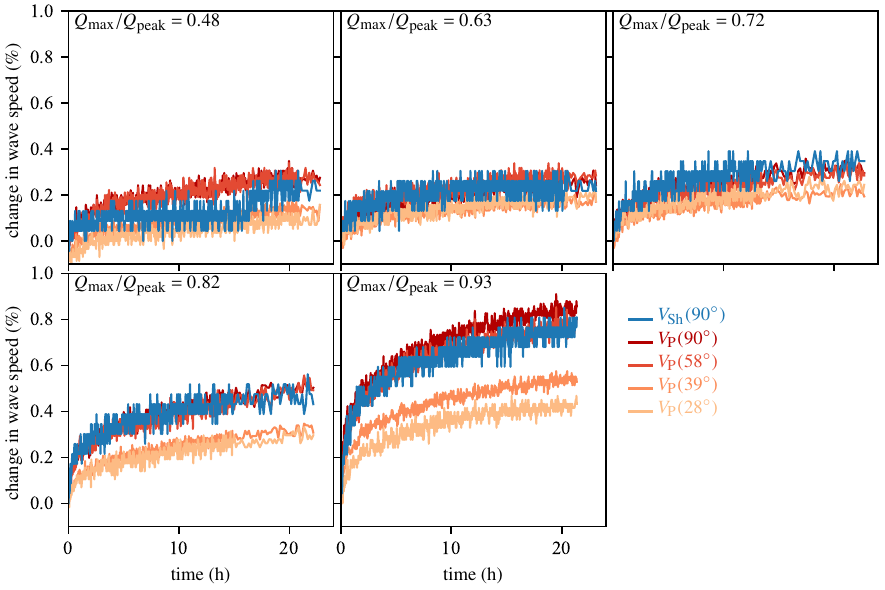}
  \caption{Relative change in elastic wave velocities as a function of time during hydrostatic hold periods after each loading cycles. Test conducted at 80~MPa confining pressure.}
  \label{fig:recoveryV}
\end{figure}

In order to investigate the long-term persistence of the residual strains and velocity drops that remain after deformation and unloading, one experiment was conducted at $P_\mathrm{c}=80$ MPa where the rock was held under hydrostatic conditions (with axial load completely removed) for 24 hours between each loading cycle. Elastic wave velocities tend to recover during the hold period, by only about 0.1 to 0.2\% after load cycles up to modest differential stress ($Q_\mathrm{max}/Q_\mathrm{peak}<0.8$), and by up to 0.8\% when the maximum stress was near the failure strength (Figure \ref{fig:recoveryV}). The recovery is logarithmic with time, and is more marked along subhorizontal orientations. The recovery in wave velocity is small (typically a fraction of a percent) in relation to the wave velocity of the intact material, but represents a large fraction of the net drop that occurred during each cycle: for instance, in the last load cycle, the horizontal P wave velocity dropped by around 110 m/s (residual after unloading), and recovered by around 50 m/s over 24h, i.e., a 45\% recovery rate.

\begin{figure}
  \centering
  \includegraphics{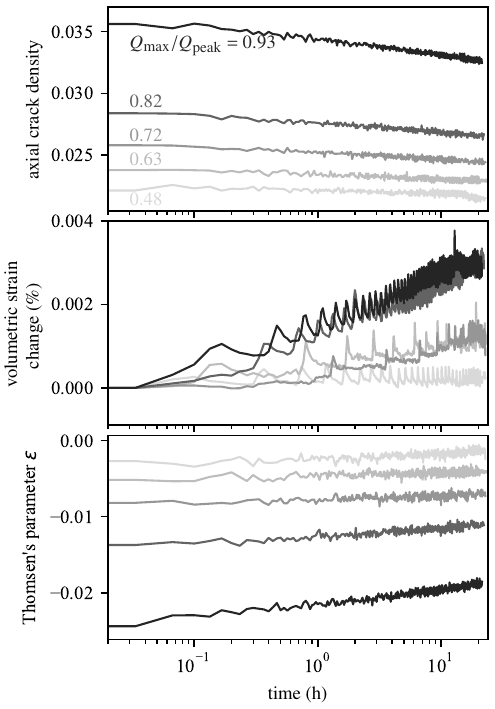}
  \caption{Evolution of vertical crack density, volumetric strain and Thomsen's $\epsilon$ parameter as a function of time during hydrostatic hold periods after each loading cycle. Test conducted at 80~MPa confining pressure.}
  \label{fig:recoveryall}
\end{figure}

The recovery in wave velocities can be interpreted as a logarithmic decrease in crack density, mostly in the axial direction (Figure \ref{fig:recoveryall}a). It is accompanied with a small but detectable logarithmic compactive change in volumetric strain, together with a decrease in elastic anisotropy (Figure \ref{fig:recoveryall}b,c). The recovery is almost negligible when the maximum differential stress was less than around 70\% of the strength of the rock: this is expected since only very small residual change in wave velocity and volumetric strain occur after those load cycles.

The recovery data are consistent with the time-dependent dilatancy recovery observed by \citet{scholz74, holcomb81} in dry granite, and with anelastic strain recovery observed by \citet{gao14} in granite, marble and sandstone. In addition, our elastic wave velocity data follow the same qualitative trends as those observed in porous limestone \citep{brantut15b} and dry Carrara marble \citep{meyer21}. 


\section{Discussion}

\subsection{Summary of results}

The main experimental observations can be summarised as follows. During deformation at relatively low stress, elastic wave velocities change only slightly, volumetric strain is reversible without hysteresis, and shear strain is largely reversible but significant hysteresis is observed. When deformation is continued up to elevated stress, typically beyond the onset of dilatancy, both volumetric and shear deformation, as well as elastic wave velocities, exhibit hysteresis. When load is removed, small residual permanent changes exist (net dilation, shear strain and velocity drop), and recover logarithmically with time under hydrostatic conditions. When the samples are re-loaded up to differential stresses lower than any previous maximum, shear strain and elastic wave velocities exhibit hysteresis, but no further permanent change is recorded and the re-loading paths all overlap. The small irrecoverable changes in elastic wave velocity (and associated anisotropy) become larger during hydrostatic decompression.

The stress-strain behaviour observed here is identical to the results of \citet{zoback75b}, and their interpretation in terms of coupled grain boundary friction and microcrack opening (a qualitative description of what became later the ``wing crack'' model, originated by \citet{brace66b}) remains largely valid in the light of complementary measurements of elastic wave velocities, including those from previous authors \citep{holcomb81,passelegue18b}. During loading, the decrease in elastic moduli is interpreted as the opening microcracks preferentially oriented along the compression axis (consistent with the anisotropic decrease in wave velocities), while inelastic axial (shear) strain may be attributed to slip along preexisting interfaces, most likely grain boundaries.

Slip along preexisting flaws is not the only mechanism that can generate tensile microcracks: stress concentrations due to open pores \citep[e.g.][]{sammis86} and elastic anisotropy or moduli mismatch at grain contacts \citep[e.g.][]{dey81} also lead to crack nucleation and growth. In Westerly granite, there is no detectable equant porosity, so pore-emanating cracks are not significant. Tensile cracking due to moduli mismatch was analysed in a simple configuration by \citet{dey81}; in this model, tensile cracking is linked to shear stresses that are proportional to differences in elastic moduli between contacting grains, but is not linked to actual slip at the contact. From a kinematic point of view, axial tensile cracks originating from this mechanism are not associated to additional axial strain and are not expected to produce hysteresis. The model of \citet{dey81} relies on a number of simplifying assumptions, and further analysis would be needed to determine how shear and tensile cracks are coupled when elastic mismatch is significant. Nevertheless, while this mechanism is potentially relevant to explain tensile cracking in our experiment on Westerly granite, it is insufficient to explain the observed hysteresis in stress-strain curves.

While the role of internal slip in the generation of tensile microcracks has been well established for several decades \citep[see review in][Section 5.8]{paterson05}, a number of features were still undocumented. In particular, the energy budget analysis shows that frictional processes largely dominate the stress-strain response, and the contribution of crack propagation becomes nonnegligible (but remains quite small) only at the highest loads. In addition, our results confirm those of \citet{holcomb81, passelegue18b} in that most of the elastic wave velocity variations are recoverable upon unloading, but also indicate the existence of small but measurable irrecoverable changes in elastic properties, volumetric strain and anisotropy. The residual changes increase with increasing maximum stress, and are amplified during hydrostatic decompression.

\subsection{Role of friction}

In order to analyse the role of friction on the deformation and physical properties of brittle rocks, it is helpful to use a micromechanical framework that contains the key necessary ingredients: slip on preexisting grain boundaries or defects and opening of tensile cracks. A number of micromechanical models of brittle deformation have been published that investigate the coupled evolution of slip and crack opening and its effect on macroscopic strain. Perhaps the first model that accounted for friction along preexisting cracks is that of \citet{mcclintock62}, who computed growth criteria for such cracks; the effect of friction has subsequently been included in many models aimed to determine \emph{thresholds} for inelastic behaviour or dilatancy \citep[][among others]{nemat-nasser82,horii85,ashby86,ashby90}. The effect of friction on \emph{stress-strain} behaviour was first analysed quantitatively by \citet{holcomb78}, with a focus on explaining dilatancy. Further work systematically investigated the coupled role of frictional slip and tensile crack growth on full stress-strain curves, using a variety of approximations to solve the crack problems and volume averaging to obtain macroscopic stress-strain curves (or tangent moduli). \citet{kachanov82a,kachanov82b} solved three-dimensional crack sliding and extension problems, and used non-interaction approximation to average the strain contributions of cracks to total strain. A similar approach with a simplified crack geometry was used by \citet{fanella88}. \citet{horii83} used a self-consistent effective medium approach to approximate crack interactions and obtain tangent moduli of rocks containing open and sliding cracks. Complete stress-strain curves were obtained by \citep{moss82,nemat-nasser88} based on a two-dimensional sliding crack model including dilation. Similar results were obtained by \citet{basista98}, who used the energy approach of \citet{rice75} to compute macroscopic strains from microscale processes. Similar energy-based micromechanical models were extended to three dimensional cases by \citet{pensee02,deshpande08,bhat11}. Overall, all these models vary in their quantitative predictions due to a number of detailed assumptions about crack geometry, kinetic relations and homogenisation procedures, but share the same qualitative features.

The model of \citet{basista98} can be chosen as an illustrative example. Although this model is two dimensional (plane strain), three dimensional effects are not expected to change its qualitative behaviour (e.g., the three dimensional model treated by \citet{kachanov82b} leads to similar predictions but scales differently with microcrack dimensions), and its application to triaxial conditions provides sufficient insights for our present purpose. Inelastic behaviour is linked to slip $b$ on preexisting cracks of length $c$, oriented with an angle $\phi$ with respect to the compression axis, and extension of axial tensile ``wings'' of length $\ell$. During triaxial loading, the increment in inelastic strain is given by \citep[][their equation (35) adjusted for stress signs]{basista98}
\begin{linenomath}
\begin{align} 
  d^\mathrm{i}\epsilon_{ij} =& \omega_0\left(\begin{array}{cc}
                                              -\sin 2\phi & \cos 2\phi\\
                                              \cos 2\phi & \sin 2\phi
                                             \end{array}\right) d(b/c) \nonumber\\
                            & + \omega_0\left(\begin{array}{cc}
                                                0 & -\sin \phi\\
                                                -\sin \phi & 2\cos \phi
                                              \end{array}\right) \left((\ell/c)d(b/c) + \frac{4(1-\nu_0^2)}{E_0}\tau_\mathrm{eff}d(\ell/c)\right) \nonumber\\
                            & - \frac{4\pi\omega_0(1-\nu_0^2)}{E_0}\left(\begin{array}{cc}
                                                                           0 & \tau_{12}/2\\
  \tau_{12}/2 & \sigma_2
                                                                         \end{array}\right)(\ell/c)d(\ell/c),      \label{eq:deij}
\end{align}
\end{linenomath}
where $\omega_0 = N_\mathrm{A}c^2$ is the two-dimensional crack density, $N_\mathrm{A}$ is the number of cracks per unit area, $\nu_0$ is Poisson's ratio of the intact material, $E_0$ is its Young's modulus, $\sigma_2$ is the applied lateral stress and $\tau_{12}$ is the applied shear stress. The stress $\tau_\mathrm{eff}$ is the effective shear stress that drives frictional slip (i.e., the shear stress in excess of the sliding resistance along the crack). Equation \ref{eq:deij} is valid in the limit where tensile ``wings'' are long (large $\ell/c$), so that they are strictly in the axial direction. The expression of inelastic strain in such a simple model contains key information to physically interpret our results. Firstly, the only contribution to inelastic axial strain is that of frictional slip (first term on the rhs). Secondly, there are coupling terms between slip and tensile crack length (second term on the rhs), which produce lateral strains. Thirdly, increments in tensile crack length do not produce axial strain, but mostly lateral strains (second and third terms on the rhs). Lastly, increments in slip and tensile crack length tend to increase the ratio of lateral to axial strains, but this is purely an \emph{inelastic} effect, i.e., linked to changes in slip and crack length. All the aforementioned effects are of course only strictly correct within the model assumptions, but yet provide the dominant, first order contributions of slip and crack growth to macroscopic strains in realistic cases.

The physical intuition provided by Equation \eqref{eq:deij} in a simple model is entirely consistent with our observation of stress-strain behaviour, which show that internal slip dominates the macroscopic response (Figures \ref{fig:strainpart}, \ref{fig:energypartall}). The amount of internal slip needed to explain the inelastic strain observed during the experiments can be roughly quantified as
$b/c \approx \epsilon^\mathrm{i}_{11}/(\omega_0 \sin2\phi)$. Using a crack density of the order of $0.2$, $\phi=\pi/4$ and a maximum inelastic axial strain of the order of $0.5$\%, the ratio $b/c$ is of around $0.025$. Using an initial shear crack size of the order of the grain size, $100$~$\mu$m, an upper bound for the finite slip at the maximum differential stress is of $2.5$~$\mu$m. Upon unloading, the residual, irrecoverable axial strain is of the order of $0.15$\% (Figure \ref{fig:residualstrain}), so that the residual internal slip is of less than $1$~$\mu$m. Such small amounts of slip are unlikely to be detectable in the microstructure, even more so than microcracks re-open during decompression and might mask small offsets between grains or along preexisting flaws.

The coupling between tensile crack size and slip is apparent in the second term of Equation \eqref{eq:deij}: tensile ``wings'' provide an additional contribution of slip to lateral strain, i.e., to dilatancy. This is apparent, for instance, in the correlation between residual volumetric strain and residual anisotropy (Figure \ref{fig:residualthomsen}): hysteresis due to friction prevents complete closure of tensile cracks, which tend to be mostly in the axial direction, so that dilatant strain evolves in tandem with anisotropy. One can also directly observe that crack growth alone is not expected to produce dilatancy without the coupling with shear stress and slip on the preexisting flaws. This feature is linked to the rather simple crack geometry of the model with purely axial tensile wings, but again shows the dominant effects.

The micromechanics of deformation arising from the wing-crack model also explains the apparent increase in Poisson's ratio at high stress, sometimes above 0.5, observed in a number of studies \citep[e.g.][]{faulkner06,heap10}. As noted in Equation \eqref{eq:deij}, the ratio of lateral to axial strain increases notably if sliding and crack growth/closure are activated. There is however a very large difference between the \emph{tangent} moduli (and the apparent Poisson's ratio that could be computed from stress-strain curves) and the \emph{elastic} moduli of the cracked rock (elastic Poisson's ratio of damaged dry rocks typically {decrease} with increasing crack density). Large Poisson's ratio can only be explained by inelastic effects (slip and crack growth/closure), which are activated by stress. In other words, large Poisson's ratio are not an inherent property of ``damaged'' rocks, but appear \emph{at high stresses} due to the activation of internal slip. The effect of ``damage'', or in a narrower sense, tensile microcracks, is to lower the stress threshold for non-linear behaviour, since the fracture toughness that gives rise to cohesion becomes zero when tensile cracks preexist. This behaviour is consistent with the ``erosion'' of the onset of dilation C' with load cycles under triaxial conditions \citep{hadley76}. Therefore, the apparent Poisson's ratio of rock depends on stress and stress history, and cannot be considered a material constant that depends solely on preexisting microstructural features.



\subsection{Characteristics of microscale friction}


During unloading, the inferred open crack density almost immediately decreases as soon as the load is reduced (Figure \ref{fig:cyclescracks}). As noted by \citet{stevens80,holcomb81}, there is apparently no ``dead band''  in a strict sense, contrary to what wing crack models would predict. However, the initial rate of crack density reduction is initially very small, and increases significantly as load is further reduced. The contrast in behaviour between successive cycles at increasing loads (Figure \ref{fig:typicalcycles}, cycles 10 and 11), where wave velocities show almost a ``true'' deadband, and those at decreasing loads (Figure \ref{fig:typicalcycles}, cycles 13 and 14) where the deadband is less apparent, is compatible with the wing crack concept. During cycles performed up to stresses lower than the previous maximum, the rock has kept the memory of the previously achieved maximum in the form of the elastic restoring force between the shear crack faces, so that reverse sliding can be activated more easily as soon as load is reduced, compared to cases where a new maximum has just been reached. Experimental observations, including those of \citet{holcomb81}, do not completely invalidate wing crack models, but rather call for a few amendments to the description of wing crack mechanics. Several non-exclusive possibilities exist to explain the near-instantaneous change in crack density upon unloading: (1) complex friction law, e.g., including direct stress dependency on friction coefficient or hysteresis, (2) viscous relaxation in the bulk (or along shear cracks or grain boundaries) and (3) elastic changes in crack aperture leading to the formation of contact points along crack faces, which have instantaneous stiffening effects on effective elastic moduli \citep{trofimov17,passelegue18b,meyer21}.

The time-dependent recovery of crack density (Figure \ref{fig:recoveryall}) also requires microscale friction to be more complex than just a constant coefficient multiplied by normal stress. Under constant, hydrostatic stress conditions, \citet{brantut15b} showed that the length of tensile wing cracks in the model of \citet{basista98} should evolve as
\begin{linenomath}\begin{equation}
  \ell(t) - \ell_0 \propto (A-B)\ln(1+t/(T(A-B))),
\end{equation}\end{linenomath}
where $\ell_0$ is the initial length immediately after unloading, $(A-B)$ is the steady-state rate-dependency of friction coefficient (assumed positive, i.e., rate strengthening), and $T$ is a time constant that depends on the initial shear crack length and geometry, confining pressure and elastic moduli of the solid.  The logarithmic recovery observed in our data (Figure \ref{fig:recoveryall}) is qualitatively compatible with this model. 

Recovery is also compatible with bulk relaxation and the generation of contact islands along cracks, as discussed in \citet{meyer21}. In the case of dry granite at room temperature, it is unlikely that viscous flow or fluid-related processes (such as pressure solution) can have any significant role in the recovery phenomenon, and backsliding (accompanied by crack closure) is most likely the dominant process.

The mechanism of backsliding requires that the frictional strength along preexisting defects such as grain boundaries does not significantly increase as the sense of shear is reversed. In our tests, under dry conditions, at relatively slow rates and with limited net slip (with no direct evidence left in the microstructure), there is no reason to believe that frictional strength can be dramatically different upon reversal of slip direction. This may not be the case in general, for example if forward slip is changing permanently the structure of the interface.

What sort of friction law would be required to explain all experimental observations? Firstly, we should recall the key observation that the dilatancy threshold increases with increasing confining pressure \citep{brace66}, which is consistent with the idea that a well-defined friction coefficient (a constant ratio between shear and normal stress at the onset of slip) exists, to first order. The near instantaneous, small backslip upon unloading inferred from the variation of wave velocities is compatible with the existence of an elastic shear stiffness of the microscale interfaces. The time-dependent recovery is compatible with rate-strengthening friction, as described (for instance) by rate-and-state laws. Therefore, friction at the microscale (grain contacts being of the order of 100$\mu$m) has all the characteristics of \emph{macroscopic} friction. Models of macroscopic friction rely on rough interfaces with elastic or plastic microcontacts \citep[e.g.][chap. 2]{scholz19}, and the microscopic frictional interfaces are therefore expected to be also rough. This is potentially due to the opening of grain boundaries due to cooling and exhumation, forming geometric mismatch between internal interfaces.

We expect microscale friction to change character at high pressure or high temperature, if all open cavities are closed: this would correspond to saturation of contact area, at which point resistance to slip is not directly proportional to normal stress but governed by interfacial defects (e.g., grain boundary dislocations) or adhesive forces. Such a transition has been inferred in experiments conducted in antigorite at high pressure and temperature \citep{david20}. The change in characteristics of microscale friction may contribute to the brittle-ductile transition, in addition to commonly recognised processes such as the activation of intracrystalline creep mechanisms.



\subsection{Contribution of tensile cracking to faulting}

Throughout this paper we have considered the evolution of average rock properties with deformation cycles. However, at the highest differential stresses reached in some of the tests, we expect strain to be at least partially localised. \citet{hadley75b} showed that circumferential strain loses circular symmetry around the compression axis at differential stress beyond 90\% of the failure stress, indicating an early tendency towards the formation of a shear fault. In situ, time-resolved x-ray tomography observations \citep{renard19} seem to indicate that strain localisation in the form of shear and tensile crack coalescence is a rather late process, occurring at stresses around 99\% of the peak. The inception of macroscopic shear faulting is caused by microcrack interactions, which facilitate further tensile crack propagation and slip on shear cracks. Despite these additional quantitative changes to the mechanics of microcracking and internal slip, interactions do not change the essential qualitative characteristics of friction and tensile crack opening at the microscale.

Damage accumulation (per inelastic strain increment) in the form of tensile microcracks is known to accelerate as macroscopic failure approaches and strain becomes localised \citep[e.g.][Section 1.2.2]{scholz02}. At this stage, the contribution of tensile cracking to the overall energy dissipation may deviate from what is observed at lower stress levels. Our data show a trend towards a more prominent contribution of open cracks at high stress (Figure \ref{fig:energypartall}). In addition, the cumulative energy dissipation from microcracks is also substantial, up to 40\% or more (if, say, the rock was brought to shear failure in a single cycle). Thus, it is possible that the energy required for tensile crack growth could transiently dominate the energy budget in the phase immediately preceding shear failure, where differential stress is near its peak.

The transition from homegeneous deformation to shear failure cannot be analysed with our laboratory methods, which require volume averaging of strains and elastic wave velocities. Nevertheless, the energy balance of pre-rupture deformation given here (Figure \ref{fig:energypartall}) is consistent with that obtained from P wave velocity tomography by \citet{aben19,aben20c} on faulted rock: during quasi-static shear failure propagation, the contribution of tensile fracturing as captured by changes in elastic moduli remains a small fraction (less than 10\%) of the total energy dissipated by other processes such as friction. More tensile crack damage is generated during dynamic faulting than in quasi-static tests, but the change is not dramatic \citep{aben20c}. The localised deformation around a shear fault (fault ``damage'' zone) is therefore of the same nature as the bulk deformation of the nominally intact rock, and is likely dominated by shear.

Tensile cracks do not need to be present in large densities or have pervasively large dimensions throughout the rock volume to produce large effects on crack coalescence, shear failure and slip. This is consistent with the fact that crack coalescence and catastrophic failure is linked to \emph{local} maxima in stress intensity factors \citep{kachanov12}, i.e., only the most deleterious cracks play a role.

Despite their small contribution to the average dissipation and finite inelastic strain during brittle rock deformation, tensile microcracks have major consequences on rock physical properties and leave unique microstructural traces, unlike shear cracks. However, like shear cracks, their impact on properties like permeability or elastic wave velocities is largely reversible and primarily stress-controlled \citep[e.g.,][]{zoback75,mitchell08,passelegue18b}. One property that is likely irreversibly impacted by tensile cracks is the onset of dilation. Following the model of \citet{nemat-nasser82} and data summarised in \citet{ashby90}, the stress at the onset of inelastic dilation is well captured by
\begin{linenomath}\begin{equation}
  \sigma_1 = \frac{\sqrt{1+\mu^2}+\mu}{\sqrt{1+\mu^2}-\mu}\sigma_3 + \frac{\sqrt{3}}{\sqrt{1+\mu^2}-\mu}\frac{K_\mathrm{Ic}}{\sqrt{\pi c}},
\end{equation}\end{linenomath}
where $\sigma_1$ and $\sigma_3$ are the axial and lateral stress, respectively, and $K_\mathrm{Ic}$ is the mode I fracture toughness of the solid. The contribution of friction provides the pressure dependency of the onset of dilation, while the resistance to mode I cracking (toughness) provides a constant offset, or cohesion. During multiple cycles past the onset of dilation, the contribution of friction remains while that of toughness disappears if cracks do not physically heal or seal. The reduction of the onset of dilatancy with stress cycles documented by \citet{hadley76}, although imprecise due to the smooth nature of stress-strain curves across the yield point, likely reflects the drop in cohesion associated with the disappearance of the toughness term.

\section{Conclusions}


A dominant component of stress-induced ``damage'' in rocks can be attributed to microscale slip on preexisting interfaces, such as grain boundaries, which is largely reversible upon unloading and does not leave any substantial microstructural record. Tensile cracking is coupled to slip, and produces large changes in dynamic elastic properties and dilatancy, but only contributes to a small fraction of the energy required for deformation. As originally discussed by \citet{holcomb81}, rocks keep a memory of the maximum differential stress applied to them. The simplest conceptual model to explain this memory effect is that the maximum stress is recorded as an elastic restoring force along frictional interfaces, well captured by wing crack models. Residual, irrecoverable strains, elastic wave velocity drops and anisotropy can be attributed to the hysteresis associated with backslip, where friction has to be overcome in the reverse slip direction. Complete decompression amplifies the small residual changes that occurred during triaxial loading.

The inelastic behaviour governed by friction and tensile fracturing implies that deformation and physical properties depend on stress path and not only on the current stress state. Due to the threshold effect of friction, ``damage'' makes rock behave in a nonlinear fashion, and is only visible at elevated stress (where slip and crack opening can be activated). In the crust, differential stress is limited by macroscopic frictional strength of faults \citep[e.g.][]{zoback92,zoback97}, which has the same characteristics as frictional strength of microscale interfaces responsible for inelastic rock behaviour (as long as the fault core composition is similar to that of the wall rock). Therefore, one expects that the manifestation of ``damage'' can only be significant in regions where local differential stress is (or has been) sufficiently large to allow for internal slip to be activated: ``damage'' has significant consequences only at elevated differential stress. If a volume of rock has been subjected to transiently high compressive shear stress, it can keep a memory of the peak stress in its elastic anisotropy, but this memory can be easily erased with further changes in stress orientation and magnitude. This has been evidenced in laboratory true triaxial experiments \citep{browning17,browning18}. In nature, we expect shear stress to fluctuate cyclically in the vincinity of seismogenic faults. Although shear stress is on average likely close to ``Byerlee''-type frictional strength, it can be locally much higher (e.g., at geometric irregularities). In addition, during earthquake propagation, near-tip stresses are also transiently well above the long-term applied stress level. Thus, rocks located close to seismogneic faults will experience stress cycles during which microscale slip and tensile cracking will occur. It is in these so-called ``damage'' zones that significant strain energy is dissipated during the seismic cycle \citep[e.g.][]{griffith10,okubo19}, and we highlight here that tensile cracking may not be the dominant contribution to this dissipation.

Time-dependent recovery in elastic wave velocities and dilatancy under constant stress conditions is consistent with rate-dependent friction along microscopic interfaces, coupled to backslip-induced crack closure. Time-dependent friction implies that rocks have a viscous behaviour even under low temperature, low pressure and dry conditions. Detailed modelling and experimental work is required to better characterise this time dependency and how it interacts with known other time-dependent aspects of brittle rock deformation, such as subcritical cracking \citep{brantut13}.  Nevertheless, the possibility of bulk stress relaxation and inelastic deformation driven by time-dependent microslip has potentially large consequences for our understanding of transient deformation phenomena in the brittle crust such as post-seismic deformation and elastic wave velocity variations during the seismic cycle.

\paragraph{Acknowledgments} Frans Aben and Emmanuel David helped running the experiments. Teng-Fong Wong and an anonymous reviewer made useful suggestions that helped clarify the paper. The code used to invert group velocity data for elastic moduli is accessible at \url{https://github.com/nbrantut/VTIModuli.jl}. The code version used here was commit \texttt{dae4376}. Support from the UK Natural Environment Research Council (grant NE/K009656/1) and from the European Research Council under the European Union's Horizon 2020 research and innovation programme (project RockDEaF, grant agreement \#804685), is gratefully acknowledged. Data are available at \url{doi:10.5281/zenodo.6761111}.

\bibliographystyle{agufull}
\bibliography{references}

\clearpage
\appendix

\renewcommand\thefigure{\thesection.\arabic{figure}}
\setcounter{figure}{0}    

\section{Alternative estimate of integrated fracture energy}
\label{ax:energy}

During a complete loading and unloading cycle, the energy dissipated is the sum of contributions including friction, crack opening and crack growth. Energy changes due to tensile crack growth can be estimated from independent measurements of elastic moduli (Section \ref{sec:energy}), but are also reflected in the overall stress-strain behaviour. Here, we establish an approximate technique to estimate an upper bound of the energy change due to tensile fracturing during a given loading cycle, based solely on stress-strain analysis.

\begin{figure}
  \centering
  \includegraphics{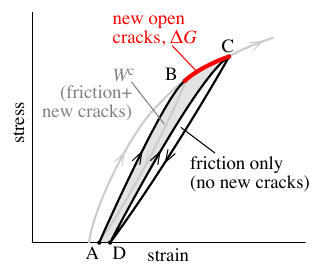}
  \caption{Sketch of stress-strain behaviour during successive loading cycles. The cycle of interest initiates in state A and ending in state D. The red portion $BC$ corresponds to the stress and strain interval where the previous maximum stress is overcome, i.e., where tensile cracks further propagate and new ones are formed.}
  \label{fig:energysketch}
\end{figure}

Let us consider a rock sample that experienced a complete loading cycle ($ABCD$ in Figure \ref{fig:energysketch}), and is now in a macroscopically unloaded state (point $D$ in Figure \ref{fig:energysketch}). Upon reloading from point D, the stress remains lower than during the previous cycle, but meets again point C, the maximum in stress and strain achieved in the previous cycle. The energy dissipated in the unloading-loading cycle $CDC$ is purely frictional, since we do not expect any significant propagation of tensile cracks as long as the applied stress remains lower than the previously achieved maximum. The difference between the total dissipated energy in the cycle (area $ABCD$) and the frictional dissipation upon reloading (area $CDC$), denoted $W^\mathrm{c}$, must therefore include all the contribution of tensile crack propagation during that cycle. It certainly includes other contributions too, such as that of forward sliding of cracks during loading from $A$ to $B$, which occurs at a higher stress state and lower tensile crack density than from $C$ to $D$. $W^\mathrm{c}$ is thus an upper bound for the contribution of tensile crack growth in a given cycle. When computing the fracture energy from changes in elastic moduli $\Delta G$ (Equation \eqref{eq:DG}), only the portion $BC$ is considered to minimise the contribution of reversible changes in moduli due to re-opening of preexisting cracks.

\begin{figure}
  \centering
  \includegraphics{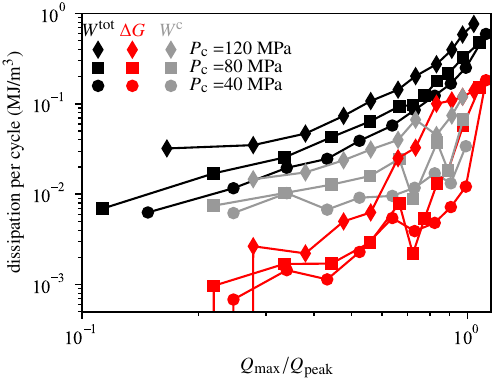}
  \caption{Comparison of total energy dissipation per cycle ($W^\mathrm{tot}$), fracture energy estimates $\Delta G$ from changes in elastic moduli and $W^\mathrm{c}$ from stress-strain analysis.}
  \label{fig:energypart2}
\end{figure}

We compute $W^\mathrm{c}$ as 
\begin{equation}
  W^\mathrm{c} = W^\mathrm{tot} - \int_{CDC}P_\mathrm{c}d\epsilon_{kk} - \int_{CDC}Qd\epsilon_\mathrm{ax},
\end{equation}
and find values that are typically much lower than the total energy per cycle $W^\mathrm{tot}$, and higher than the independent estimate of fracture energy $\Delta G$ (Equation \ref{eq:DG}; Figure \ref{fig:energypart2}). In absence of elastic wave velocity data, using $W^\mathrm{c}$ as a proxy for fracture energy would lead to significant overestimations (by factors of 2 to 8), but would still lead to the correct conclusion that the energy required to propagate tensile cracks is around one order of magnitude smaller than the total dissipated energy.

\end{document}